\documentclass[aps,prd,reprint,preprintnumbers,superscriptaddress,nofootinbib,longbibliography]{revtex4-2}
\usepackage[utf8]{inputenc}
\usepackage{amsmath,amssymb,array,calc,rotating,epsfig,psfrag}
\usepackage{braket}
\usepackage{natbib,hyperref}
\usepackage[capitalise]{cleveref}
\usepackage{float}
\usepackage{tensor}
\usepackage{svg}
\usepackage{xspace}
\usepackage{lineno}

\newcommand{\tr}{\text{tr}}
\newcommand{\pythia}{\textsc{Pythia}\xspace}

\begin{document}
\preprint{FERMILAB-PUB-22-912-SCD}

\title{Prolegomena to a hybrid classical/Rydberg simulator for hadronization (\textsc{QuPyth})}

\author{Blake Senseman}
\affiliation{The University of Iowa, Iowa City, IA 52242, USA}
\author{Zane Ozzello}
\affiliation{The University of Iowa, Iowa City, IA 52242, USA}
\author{Kenneth Heitritter}
\affiliation{The University of Iowa, Iowa City, IA 52242, USA}
\affiliation{Fermi National Accelerator Laboratory, Batavia, IL 60510, USA} 
\author{Yannick Meurice}
\affiliation{The University of Iowa, Iowa City, IA 52242, USA}
\author{Stephen Mrenna}
\affiliation{Fermi National Accelerator Laboratory, Batavia, IL 60510, USA}

\date{\today}

\begin{abstract}
    Programmable neutral-atom arrays provide a promising route to real-time analog simulation of strongly interacting quantum systems. We introduce a two-leg Rydberg-atom ladder that realizes string dynamics and controllable particle production using experimentally accessible parameters. A mapping between local Rydberg occupations and an emergent electric field yields charge–anticharge pairs connected by dynamical strings. Classical simulations enforcing Rydberg blockade constraints identify regimes with suppressed entanglement spreading and tunable particle multiplicities, which are seen to be signatures of confinement and string breaking. Particle multiplicities typically grow monotonically with time and system size and depend sensitively on simulator detuning and interaction scales. These results establish the ladder geometry as a viable near-term analog quantum simulator of string fragmentation, and motivate hybrid workflows in which quantum devices contribute nonperturbative real-time dynamics to event generation.
\end{abstract}

\maketitle
\section{Introduction}
Programmable arrays of neutral atoms with highly excited Rydberg states present a significant new tool to study quantum many-body systems \cite{Bernien2017,barredoSyntheticThreedimensionalAtomic2018,keeslingQuantumKibbleZurek2019,browaeysManybodyPhysicsIndividually2020,schollQuantumSimulation2D2021,ebadiQuantumPhasesMatter2021,bluvsteinQuantumProcessorBased2022}. More generally, the idea of using analog platforms, like those constructed of Rydberg atoms, to simulate quantum systems has received a great deal of attention \cite{blochQuantumSimulationsUltracold2012a, lewensteinUltracoldAtomicGases2007, ciracColdAtomSimulation2010, kapitOpticallatticeHamiltoniansRelativistic2011,kunoQuantumSimulationDimensional2017, danshitaCreatingProbingSachdevYeKitaev2017, ploop, davoudiSimulatingQuantumField2021,davoudiAnalogQuantumSimulations2020, monroeProgrammableQuantumSimulations2021,gonzalez-cuadraQuantumSimulationAbelianHiggs2017,aidelsburgerColdAtomsMeet2022,schweizerFloquetApproachMathbbZ2019, meuriceDynamicalGaugeFields2011}. Recent work has found these Rydberg arrays can realize gauge theories in the form of the spin-\textonehalf\xspace truncated 1+1D Schwinger model \cite{Surace2020}, spin-\textonehalf\xspace 2+1D lattice gauge theories \cite{celiEmergingTwoDimensionalGauge2020}, and for the spin-1 Abelian-Higgs model \cite{cara}. A variety of spin chains and lattice gauge theories in 1+1D and 2+1D \cite{Vovrosh2021,tanDomainwallConfinementDynamics2021,pichlerRealTimeDynamicsLattice2016,kuhnNonAbelianStringBreaking2015,Banerjee2012, hebenstreitRealTimeDynamicsString2013,kasperSchwingerPairProduction2016, verdelRealtimeDynamicsString2020, buyensRealtimeSimulationSchwinger2017, salaVariationalStudySU2018, spitzSchwingerPairProduction2019, parkGlassyDynamicsQuark2019, magnificoRealTimeDynamics2020, notarnicolaRealtimedynamicsQuantumSimulation2020, chandaConfinementLackThermalization2020, pardo_resource-efficient_2023,Cochran2025,GonzalezCuadra2025,De2024,Ciavarella2024} have been shown to display dynamical features related to QCD and the phenomenon of string-breaking. 

The production of strongly-interacting particles (hadrons) from quarks and gluons is known as hadronization,  and is understood qualitatively from the confining properties of QCD.   
Practical, first-principle predictions for hadronization are not currently available, and phenomenological models are used as a proxy when comparing field theory calculations to data.   One such model, the Lund string model \cite{anderssonPartonFragmentationString1983}, is implemented in the widely-used high-energy event generator \pythia \cite{bierlichComprehensiveGuidePhysics2022, sjostrandPYTHIAPhysicsManual2006a}.
The development of hadronization models has been identified as an avenue of great practical relevance for the continued exploration of the Standard Model \cite{campbellEventGeneratorsHighEnergy2022}.


In the Lund string model, the effective linear, confining potential between color charges gives rise to a color flux tube.  
To a first approximation, 
the tube is a massless, relativistic string
with negligible transverse dimensions.
Hadron spectroscopy indicates that such a string has
energy per unit length 
$\kappa \approx 1$ GeV/fm. 

 \begin{figure}[b!]
    \centering
    \includegraphics[width=0.9\linewidth, height=0.9\linewidth]{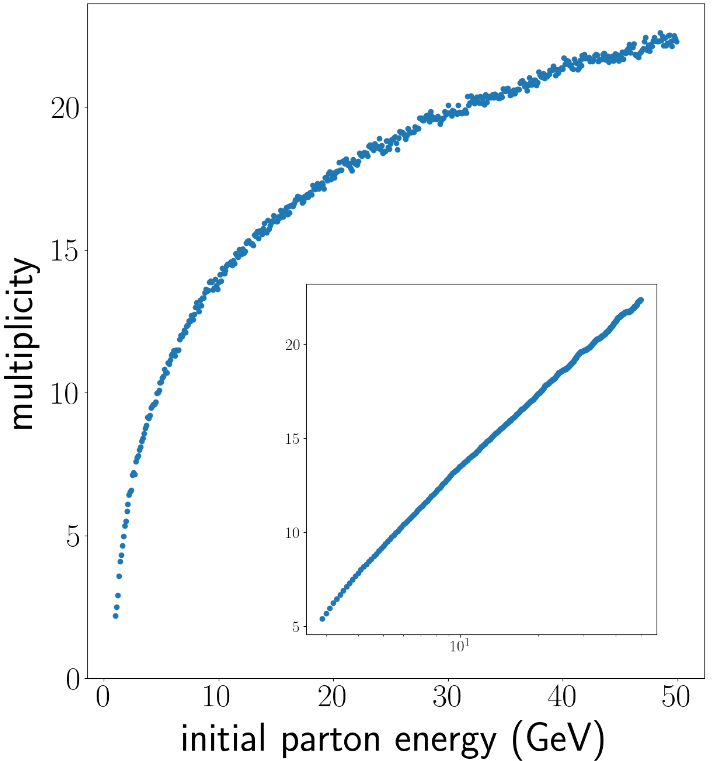}
    \caption{Hadron multiplicity predicted by \pythia from an initial $u\bar{u}$ pair.   The quantitative results depend upon several parameters fit to data.   A primary goal for a hadronization model based on Rydberg atoms is to achieve a similar logarithmic behavior. The inset displays the same data with a logarithmic x-axis and smoothing applied.}
    \label{fig:pythia_mult}
\end{figure}

As color-anticolor charges separate in spacetime,
the potential energy stored in the string
increases until it can produce a new quark-antiquark pair from vacuum fluctuations, splitting the original string into a string and a hadron.  This breaking continues until the entire string is converted into hadrons.  The  average number of hadrons (multiplicity) produced from string hadronization predicted by \pythia is shown in \cref{fig:pythia_mult}.

In this article, we discuss the possibility of replacing the 1+1 dimensional Lund model by a 1+1 dimensional quantum lattice model, namely the Abelian Higgs model in the spin-1 approximation and to use the associated ladder Rydberg simulator \cite{cara} as an event generation module that can be integrated into the existing form of \pythia. The Abelian Higgs model is not expected to match QCD exactly, and the closeness with the Rydberg simulator is not essential. However, it is important to make sure that the Hilbert spaces allow the interpretation of the latter as an approximate implementation of the former. The main challenge that we address in this article is to show that using the simulator, it is possible to prepare entangled quantum states with tunable energy and multiplicity in a way inspired by \cref{fig:pythia_mult}.

The article is organized as follows. Section II defines an effective Hamiltonian of a ladder of Rydberg atoms based on the spin-1 Abelian-Higgs model. Section III describes the measurement within this model of observables previously correlated with confinement and string breaking. These observations lead us to conclude that this is a compelling arrangement to study hadronization in a quantum model, and look forward to the development of models that can imitate classical event generation schemes.

\section{Rydberg Ladder Model}
The dynamics of Rydberg atom configurations are governed by the Hamiltonian
\begin{align}
\label{e:Rydberg_H}
H=&\frac{\Omega}{2}\sum_{j} \left(\ket{g_j}\bra{r_j} + \ket{r_j}\bra{g_j}\right) -\Delta \sum_{j} n_j\nonumber\\
&+\sum_{j<k} V_{jk}n_j n_k\;.
\end{align}

The Rydberg atoms are qubits with ground states $\ket{g}$ and excited (Rydberg) states $\ket{r}$. These states are coupled by the first term in the Hamiltonian, representing Rabi flipping with Rabi frequency $\Omega$. The second term represents the global laser detuning, which varies the energy gap between the ground state $\ket{g}$ and excited state $\ket{r}$. Here, the Rydberg density operator $n_j\equiv\ket{r_j}\bra{r_j}$, and the parameter $\Delta$ is called simply the detuning. The last term is the repulsive (van der Waals) interaction between excited atoms $\ket{r}$ with potential given by

\begin{align}
\label{e:ryd_interaction}
    V_{jk}=\frac{C_6}{|r_j-r_k|^6}\;,
\end{align}
where $C_6$ is the Rydberg interaction constant particular to the specific Rydberg state used and $|r_j-r_k|$ is the distance between atom $j$ and $k$.

\begin{figure}[b!]
    \centering
    \includegraphics[scale=0.22]{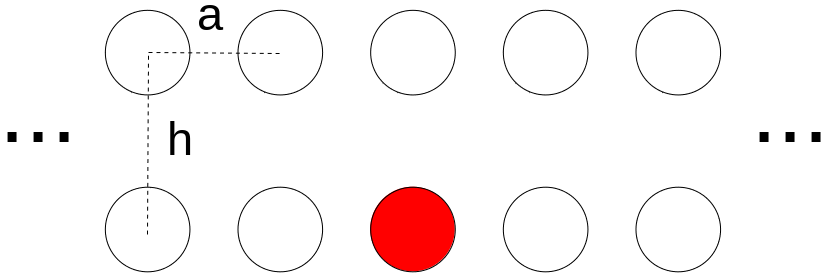}
    \caption{Two-leg ladder arrangement of Rydberg atoms where $h$ is the on-rung (vertical) spacing and $a$ is the inter-rung (horizontal) spacing. The ladder can be defined by an inverse aspect ratio $\rho=h/a$, and we work with $\rho=2$, whereas previous studies \cite{cara} have focused on $\rho \sim 0.4$. The red circle represents an atom in the Rydberg state, while the white circles correspond to ground states.}
    \label{fig:ladder_single}
\end{figure}

In this work, we consider Rubidium-87, so that $\ket{r}=\ket{70 S_{1/2}}$ and $C_6=862690\times 2\pi\;\text{MHz}\cdot \mu\text{m}^6$. The strong repulsive interaction \cref{e:ryd_interaction} causes simultaneous excitations between sufficiently close atoms to become energetically prohibited. The radius at which this excitation-blocking effect overcomes the Rabi flipping is known as the Rydberg blockade radius and is defined to be the distance where
\begin{align}
    \label{e:blockade_radius}
    \frac{C_6}{R_b^6}=\Omega\;,
\end{align}
such that $R_b=\left(C_6/\Omega\right)^{1/6}$. The choices of parameters will be discussed in Section II-B.

\begin{figure}[t!]
    \centering
    \includegraphics[scale=0.2]{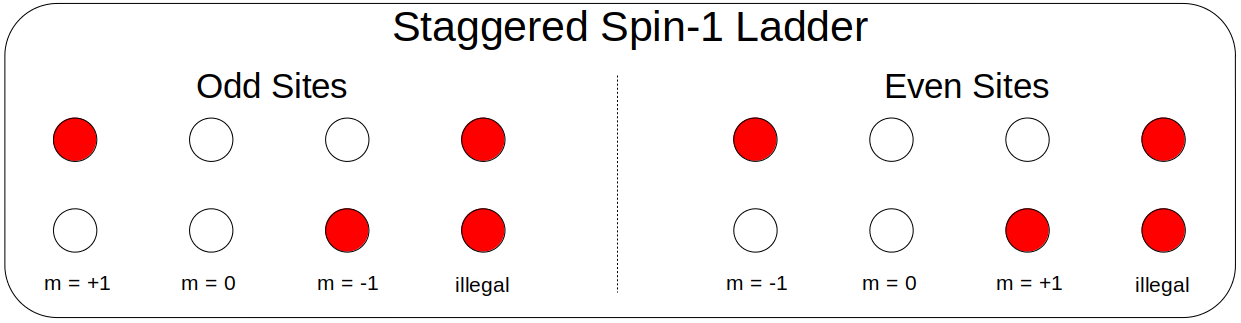}
    \caption{State mapping for odd and even sites of a two-leg ladder. Red (white) circles denote Rydberg (ground) states.}
    \label{fig:staggered_spin_one_ladder}
\end{figure}
\subsection{Two-leg ladder construction}

We study a 2$\times N_s$ arrangement of Rydberg atoms that form a two-leg ladder as displayed in \cref{fig:ladder_single}. 
Recently, a set of rung-local operators on this ladder arrangement has been shown to have a suggestive relationship with a spin-1 truncation of the Abelian-Higgs model for a specific set of parameters \cite{cara, zhangEffectiveHamiltonians}. Since this continuum theory features confinement of its scalar matter in 1+1D, we take this class of Rydberg atom models as a good opportunity to search for observables that suggest confinement and string breaking.

Following the convention of \cite{zhangEffectiveHamiltonians}, a staggered electric field operator is defined from the Rydberg densities $n_{j,k}$, labelled with rung (j $\in$ [1,..., $N_s$]) and leg (k $\in$ [0, 1]) indices.
\begin{align}
    E_j\equiv \left(-1\right)^{j+1}\left(n_{j,1}-n_{j,0}\right) = \hat{L}^z_j
\end{align}
The parity staggering (diagrammed in Fig. \ref{fig:staggered_spin_one_ladder}) is chosen so that the naturally antiferromagnetic Rydberg interactions induce ferromagnetic interactions between the electric field defined on adjacent rungs.
This can be seen from the effective Hamiltonian \cite{zhangEffectiveHamiltonians}
\begin{align}
\hat{H}^{\mathrm{eff}}_{2\mathrm{LR}}
= &\frac{\Omega}{2} \sum_{i=1}^{N_s}
\left( \hat{U}^+_i + \hat{U}^-_i \right)-\Delta \sum_{i=1}^{N_s} \left( \hat{L}^z_i \right)^2\nonumber\\
&+\sum_{k}
\bigg(
\frac{V^{(k)}_2 - V^{(k)}_1}{2}
\sum_{i=1}^{N_s - k}
\hat{L}^z_i \hat{L}^z_{i+k}
\nonumber\\
&+\frac{V^{(k)}_1 + V^{(k)}_2}{2}
\sum_{i=1}^{N_s - k}
\left( \hat{L}^z_i \right)^2
\left( \hat{L}^z_{i+k} \right)^2\bigg)
\end{align}

\newpage
\onecolumngrid

\begin{figure}[t!]
    \centering
    \includegraphics[scale=0.16]{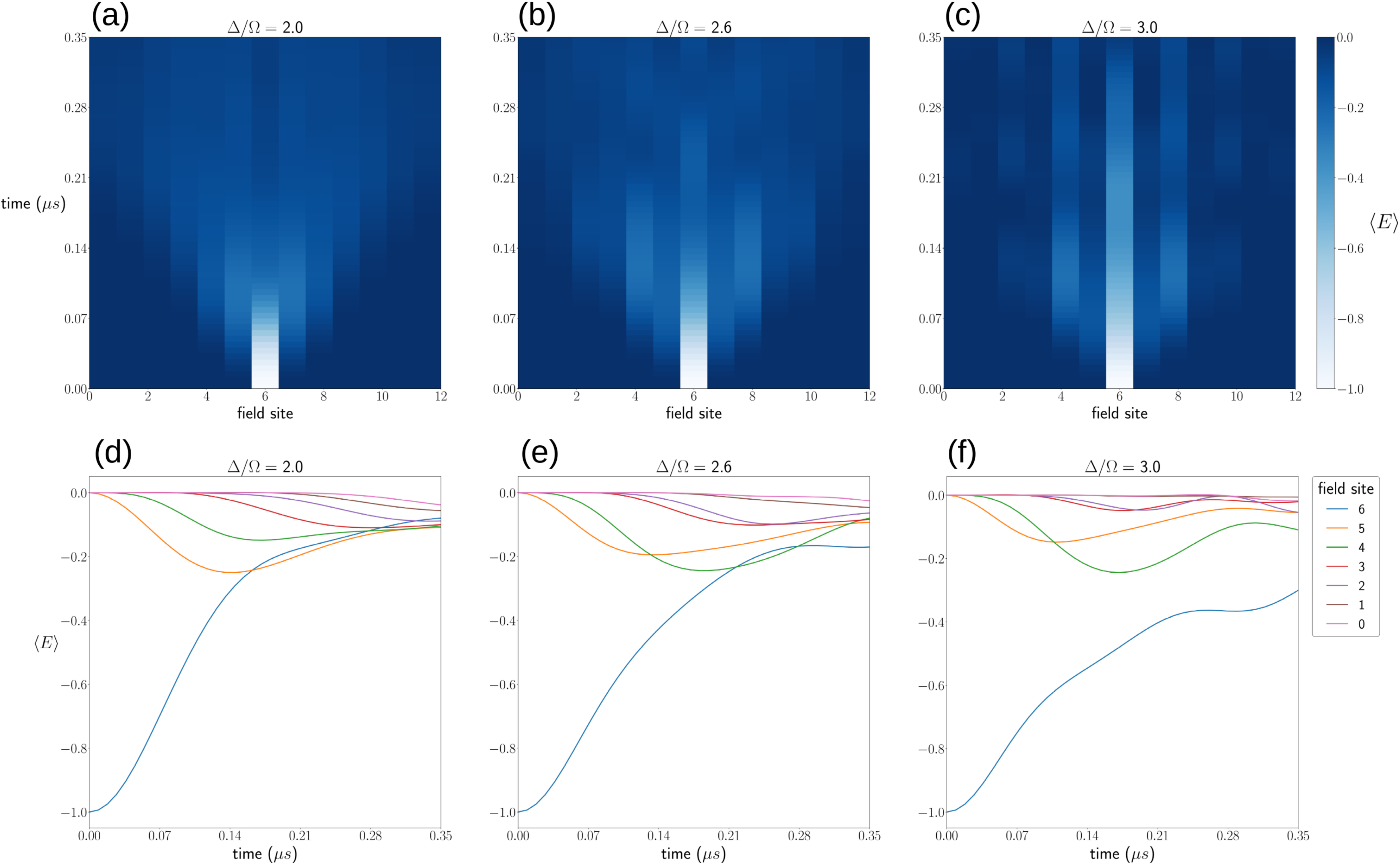}
    \caption{Each column displays the field evolution (a-c) and field evolution of individual sites (d-f) for $R_b/a=2.173$ and variable $\Delta/\Omega$ increasing toward the right. For $\Delta/\Omega=2.0$ (a), the region containing non-zero E field grows at a constant rate, suggesting a ballistic (constant momentum) propagation of local properties. As $\Delta/\Omega$ is increased, the initial field retains its density at or close to the central site, reducing the spread and increasing field localization.} It may be possible to interpret the initial state as having a higher (lower) energy for lower (higher) $\Delta/\Omega$. In this case, the initial field spreads easily when started with higher energy, but spreading is suppressed for a lower energy. Note that the field evolution in (a-c) depicts the average electric field and is, therefore, not exactly interpretable in terms of hadron multiplicity.
    \label{fig:evolution_grid}
\end{figure}
\twocolumngrid

\noindent where $V_1^{(k)} \equiv C_6/(ak)^6$ and $V_2^{(k)} \equiv V_1^{(1)}/(k^2+\rho^2)^3$ are the on-leg and cross-leg Rydberg interactions at a distance of $k$ rungs, and $U^\pm$ are the spin-1 ladder operators.

A charge representation can be defined via the application of Gauss' law
\begin{align}
\label{e:Gauss'_law}
Q_{i,i+1}=E_{i+1}-E_i\;,
\end{align}
where we refer to $Q=\pm 1$ as charge/anticharge, respectively. In the staggered field representation, the Rydberg blockade can be seen to prevent $E$-field configurations like (...,+1,-1, ...) giving rise to $Q>1$, because these states have excited atoms as nearest-neighbors. The Gauss law definition of charge also means that each pair of charge/anticharge is separated by a region of constant electric field valued $\pm1$. These regions will serve as the strings for our model. 

Using operator identities, the site non-local part of the effective Hamiltonian can also be written

\begin{align}
\hat{H}^{\mathrm{eff}}_{\text{int}}
= &\sum_{k}
\bigg(
-V^{(k)}_1
\sum_{i=1}^{N_s - k}
\hat{L}^z_i \hat{L}^z_{i+k}
\nonumber\\
&+(V^{(k)}_1 + V^{(k)}_2)
\sum_{i=1}^{N_s - k}
\mathcal{P}_{i,i+k}\bigg)
\end{align}
with
\begin{align}
    \mathcal{P}_{i,j} &\equiv \ket{1}\bra{1}_i\otimes\ket{1}\bra{1}_j \nonumber\\
&+\ket{-1}\bra{-1}_i\otimes\ket{-1}\bra{-1}_j.
\end{align}
which is an operator known to photonic models as a cross-Kerr term \cite{Malekakhlagh2020CrossResonance, Blok2021Scrambling}. This form makes more evident the antiferromagnetic ordering of adjacent sites and isolates $\mathcal{P}_{i,j}$ as the operator that spoils an exact matching with the truncated Abelian-Higgs model. Generally speaking, this term will induce non-linearity in the string potential and a weak repulsion between nearby separated strings, but these features are not a \textit{prima facie} obstacle to an event generation procedure.

In representing this effective Hamiltonian on the ladder, the Rydberg repulsion serves a dual purpose: 1) dynamically preventing two Rydberg states on a rung to preserve the effective spin-1 subspace, and 2) suppressing multiplicity with the ferromagnetic interaction term (first term in equation 7). This means that encouraging hadronization requires permitting some possibility of a state that we cannot interpret according to our field description. Future work in varying $\rho$ could investigate its effect on this tradeoff. The following section will discuss our approach to this practical challenge in the present study.



\subsection{Numerical simulation}
We use \textit{Bloqade}, a Julia package developed by QuEra to prepare the Rydberg atom configuration, device Hamiltonian, and conduct a real-time evolution. QuEra's Rydberg atom device, Aquila, has hardware limits for the various model parameters and atom geometry, which we set out to respect in our numerical simulation. $\Omega$ must be $\leq 5\pi$ MHz and the distance between atoms must be $\geq 4 \mu m$. We chose to use the minimum lattice spacing to afford the greatest expressiveness of the model.

The choice of $\Omega$ is informed by the roles the Rydberg blockade plays in the model. The effective electric field defined on each rung has no interpretation of $\ket{rr}$, so this state needs to be dynamically suppressed with $R_b/(2a) > 1$. We found $\Omega= 4\pi$ MHz to be a reasonable maximum, yielding $R_b/(2a) = 1.086 \; (R_b/a=2.173)$. Decreasing $\Omega$ or $\rho$ strengthens the blockade. To aid in computing time and exclude $\ket{rr}$ states at runtime, our numerical evolution is performed on a reduced Hilbert space that enforces the Rydberg blockade as a hard constraint. Longer range Rydberg repulsions are retained as usual in the simulator Hamiltonian. On a real device, $\ket{rr}$ can be excluded by post-selection. This choice of $\Omega$ also strongly prevents oppositely aligned E-fields on nearest neighbor sites, which corresponds to a charge of $\pm 2$ between the field sites. This is not necessary for the well-behavior of the simulator, but is a helpful feature for analyzing bitstring results at a glance.

The initial state has a single $E=-1$ site in its center, as
shown in \cref{fig:ladder_single}, which corresponds to a minimum size hadron with a charge and anti-charge on its left and right sides respectively. This state can be prepared by starting with the ladder with all atoms in the ground state and adiabatically applying a local detuning to a single central atom to flip it, before quenching to the dynamical parameters without local detuning. Aquila, a publicly available Rydberg atom device maintained by QuEra Computing Inc. \cite{Aquila_site}, can implement these preparations, the feasibility of which for this purpose will be investigated in future work.

While not all domains of the parameter space in this model will be suitable for modeling hadronization, this work begins with a broad sweep of the available configurations with the aim of implementing observations correlated in the literature with confinement and string breaking. Following the spirit of event generators, we hope these observations will be a guide to interesting parts of the model.

\section{Event Generation Features}

Previous studies of confinement on quantum devices for spin-chains \cite{Vovrosh2021,verdelRealtimeDynamicsString2020} and gauge theories \cite{Surace2020,pichlerRealTimeDynamicsLattice2016} have usually begun with extensive discussion of quasiparticle dynamics, but we will not do so yet in these Rydberg atom models. Instead, we focus on the time-evolution of field values, entropy spreading, energy-like observables, and stationary multiplicities since previous studies have found these observables suggestive in recognizing confinement and hadronization. Future work will identify the effective meson spectrum and discuss lattice effects like the Stark localisation, as has been done thoroughly for qualitatively similar models \cite{suraceScatteringMesons2021}.
Our results here provide credence to the claim that ladder configurations of Rydberg atoms present a new avenue to study confinement and may find applications to study real-time string-breaking.

\subsection{Field evolution}  Examples of the evolution for various parameters are displayed in \cref{fig:evolution_grid}, where the spin-1 staggered mapping has been implemented. The simulated evolution of $\langle E_i \rangle$ in \cref{fig:evolution_grid} shows how the initial 100\% probability central $E=-1$ field propagates outward symmetrically with a constant slope that is independent of $\Delta/\Omega$, which controls the energy density of the field. In the charge representation, this demonstrates coherent, constant velocity average motion of charges/anticharges in opposite directions. This motion appears relatively unrestricted for lower $\Delta/\Omega$ and generally becomes more constricted as $\Delta/\Omega$ is increased. 

\begin{figure}[t!]
    \centering
    \includegraphics[scale=0.18]{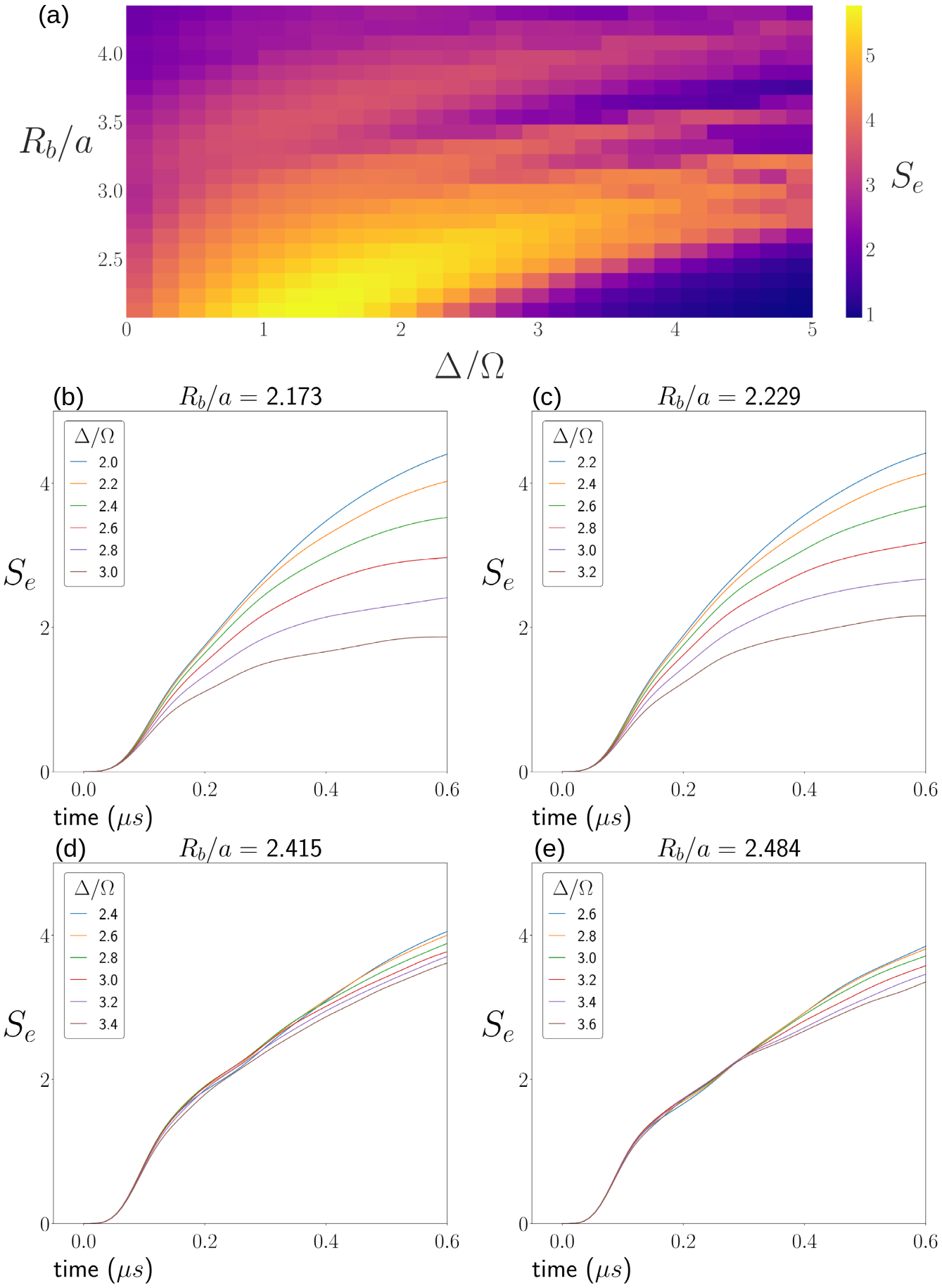}
    \caption{(a) Maximum half-chain von Neumann entanglement entropy of a 13-rung ladder after evolving for $\Delta t=1.8 \; \mu s$. There is a maximum toward the bottom-left corner of the diagram and noticeable minima structure as one moves upward along the right side of the phase-space. These minima are likely due to the presence of $Z_2$, $Z_3$, and $Z_4$ ordered phases known to be present in the ground state of the Rydberg configuration \cite{keeslingQuantumKibbleZurek2019}. (b-e) Real-time entropy evolution for a small range of parameters. All entropy evolutions approach asymptotic values at large times ($1.8\mu s$). (b) and (c) display clear suppression of entropy spreading as $\Delta/\Omega$ is increased. Previous works have pointed toward this effect as a signature of confinement.}
    \label{fig:max_entropy}
\end{figure}
\subsection{Entropy evolution}
It has been suggested that one prominent observable of confinement is a reduction in the real-time spread of entropy. This behavior was first shown in terms of a generic CFT and numerically for the XY chain with a transverse magnetic field \cite{calabreseEvolutionEntanglementEntropy2005}. An analytic result was thereafter produced for the XY chain \cite{fagottiEvolutionEntanglementEntropy2008}. Since then, numerical studies of the Ising model with a transverse and longitudinal magnetic field \cite{kormosRealtimeConfinementFollowing2017}, as well as implementation of the same model on a digital quantum computer \cite{Vovrosh2021}, have continued to build evidence relating suppression of entropy spreading to confinement. 

In all these cases, the tendency for entanglement entropy growth is correlated with the multiplication of the confined quasiparticles. We take a similar approach to substantiate our claim that the two-leg ladder on a Rydberg atom architecture may host a confining model in certain regions of its parameter space. Calculation of entropy on a quantum computer is notoriously difficult, since it generally requires some form of state tomography. Recent studies have developed new experimental methods, using two-copy interferometry, that can reliably measure entropy for Rydberg atom systems \cite{islamMeasuringEntanglementEntropy2015, bluvsteinQuantumProcessorBased2022}. This opens a path to experimental verification of our entropy observations, which will be explored in future work.

We define the half-chain reduced density matrix $\rho_{\frac{1}{2}} = \tr_{\frac{1}{2}}(\rho)$, where the total density matrix is traced over half (horizontal direction) of the atomic ladder, excluding the central rung. Inclusion of the central rung yields the same value, up to minute numerical effects, because of the symmetry of both the ladder and the definition of entanglement entropy. From the half-chain density matrix, we define the von Neumann entanglement entropy $S_{e}=-\tr\left(\rho_{\frac{1}{2}}\ln \rho_{\frac{1}{2}}\right)$. Selecting the maximum entropy over a finite time-evolution, we perform a broad sweep of the Rydberg evolution parameters in \cref{fig:max_entropy} which displays parameter dependence similar to phase diagrams of the Rydberg ladder \cite{keeslingQuantumKibbleZurek2019}. We single out examples within this phase diagram which displays reduction in entropy spreading as a function of $\Delta/\Omega$.

\begin{figure}[b]
    \centering
    \includegraphics[width=0.9\linewidth, height=0.7\linewidth]{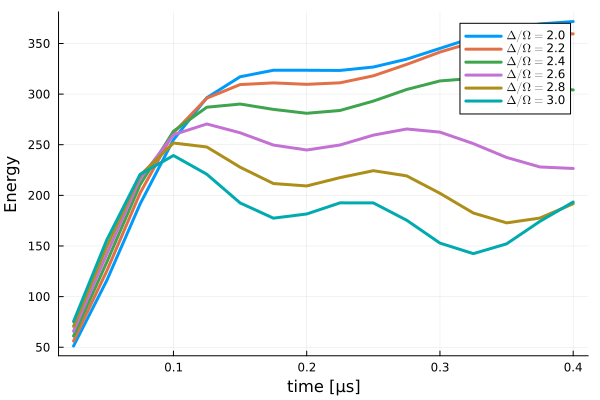}
    \caption{$\braket{H_m}$ along the evolution representing a state preparation process} 
    \label{fig:energy_evol}
\end{figure}

\subsection{Energy evolution} 
To ensure that the single-site energy of $E=\pm1$ states is higher than $E=0$ requires that $\Delta$ take a negative value. This suggests that hadronization models which more closely mirror physical confinement will be found in that region of the parameter space. However, much of the interesting evolution behavior of this model has also been observed in the $\Delta>0$ regime. While these evolutions are not suitable for target real-time evolution because of their repulsive interactions, they are valuable in preparing and manipulating states. While evolving under a Hamiltonian $\hat{H}(\Omega_0,\Delta_0)$ with $\Delta_0>0$, we measure the physical energy of the state using 
\begin{equation}
    \hat{H}_m \equiv \hat{H}(\Omega_0,-\Delta_0)
\end{equation}
This naturally will not be conserved throughout the evolution.

The curves in \cref{fig:energy_evol}
demonstrate that some combinations of parameters pump energy into the state from the perspective of the confined parameters, making them suitable regimes for state preparation, followed by a quench to the confining regime. Such prepared states would generically be coherent, multi-particle quantum states in contrast to the simple Lund string starting point.


\subsection{Multiplicity evolution}
After the evolution for a certain period of time, the state can be measured in the Rydberg density basis $\{n_i\}$ and the number of charge/anticharge pairs separated by a constant $E=\pm1$ field value can be counted. We call this count the multiplicity as it relates to the stationary ($\Omega=0$) count of hadrons.

\begin{figure}[h!]
    \centering
    \includegraphics[width=0.9\linewidth, height=0.7\linewidth]{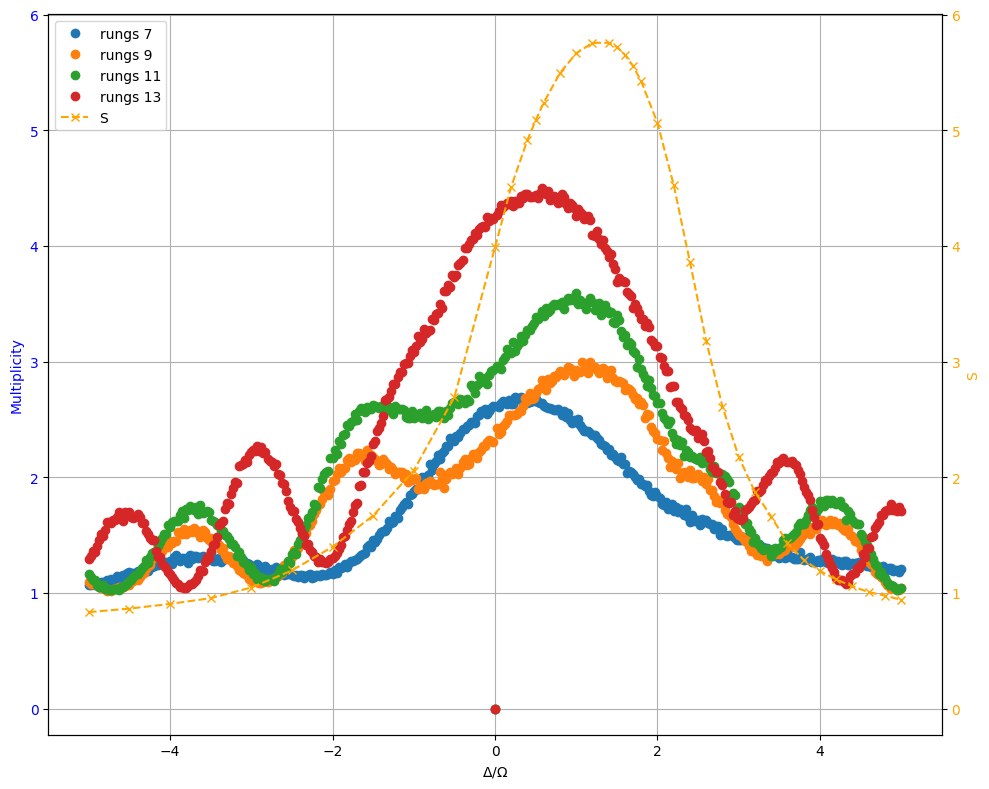}
    \caption{Average hadron multiplicity on different simulator sizes for $R_b/a=2.173$ after evolving for $\Delta t = 0.35 \; \mu s$. The increased multiplicity for smaller values of $\vert \Delta/\Omega\vert$ indicates its role in suppressing the proliferation of hadrons. Additionally, the yellow dashed line represents the half-chain entanglement entropy for 13 rungs.} 
    \label{fig:hadrons}
\end{figure}

The multiplicity of the initial state is 1 and its expectation value is generally non-decreasing throughout the evolution. The maximum capacity of the simulator is $(N_s+1)/2$, which is roughly half-saturated at most in the cases we investigate here. The curves in \cref{fig:hadrons} also illustrate periodic values of $\Delta / \Omega$ away from  0 where the increase in multiplicity is almost completely suppressed. The central bump across different numbers of rungs roughly colocates with the maximum proliferation of entanglement entropy, illustrating the connection between the motion of non-zero field values and the spread of correlations noted in other works \cite{calabreseEvolutionEntanglementEntropy2005}.

\section{Conclusion} 
We have shown that a two-leg ladder configuration of Rydberg atoms displays increases in entanglement, multiplicity, and model energy when interpreted according to a 1+1D effective field model for values of $R_b/a \gtrapprox 2$. The correlated changes in these quantities illustrate the preparation of entangled multi-particle states suitable for hadronization analysis. This is in line with previous studies that have found these observations to be signatures of confinement in a quantum lattice model. Studies of longer ladders at different aspect ratios and with a variety of state preparation methods will likely be capable of producing more realistic multiplicities, but our observations suffice to motivate the addition of the Rydberg ladder model to a growing list of quantum lattice models in which string-breaking can be studied. Future studies may also be capable of probing the string tension associated with our model via interference experiments investigated in \cite{zoharTopologicalWilsonloopArea2013, zoharSimulatingDimensionalLattice2013a}. Another potentially fruitful avenue is the analysis of the internal structure of the mesons \cite{suraceScatteringMesons2021} and measurement of their masses via entropy oscillations, as originally demonstrated in \cite{kormosRealtimeConfinementFollowing2017}. We hope that these results inspire others to further develop 
real-time simulations of QCD string-breaking, as advances in quantum hardware make such simulations feasible.

\begin{acknowledgements}
K.H. acknowledges support from the URA Visiting Scholars Program. K.H. and Y.M. are supported in part by the Dept. of Energy under Award Number DE-SC0019139. 
S.M. is supported in part by the U.S. Department of Energy, Office of Science, Office of High Energy Physics QuantISED program under the grants ``HEP Machine Learning and Optimization Go Quantum'', Award Number 0000240323, and ``DOE QuantiSED Consortium QCCFP-QMLQCF'', Award Number DE-SC0019219. This manuscript has been authored by Fermi Research Alliance, LLC under Contract No. DEAC02-07CH11359 with the U.S. Department of Energy, Office of Science, Office of High Energy Physics. This research was supported in part through computational resources provided by The University of Iowa, Iowa City, Iowa.

We thank Erez Zohar, Johannes Knaute, Johannes Zeiher, Debasish Banerjee, Fangli Liu, members of the QuLAT collaboration, and QuEra Computing Inc. for valuable comments and discussions. 

\end{acknowledgements}

\appendix

\section{Time Evolution of Entanglement Entropy and Multiplicity}
\begin{figure}[t!]
    \centering
    \includegraphics[width=1\linewidth, height=1.25\linewidth]{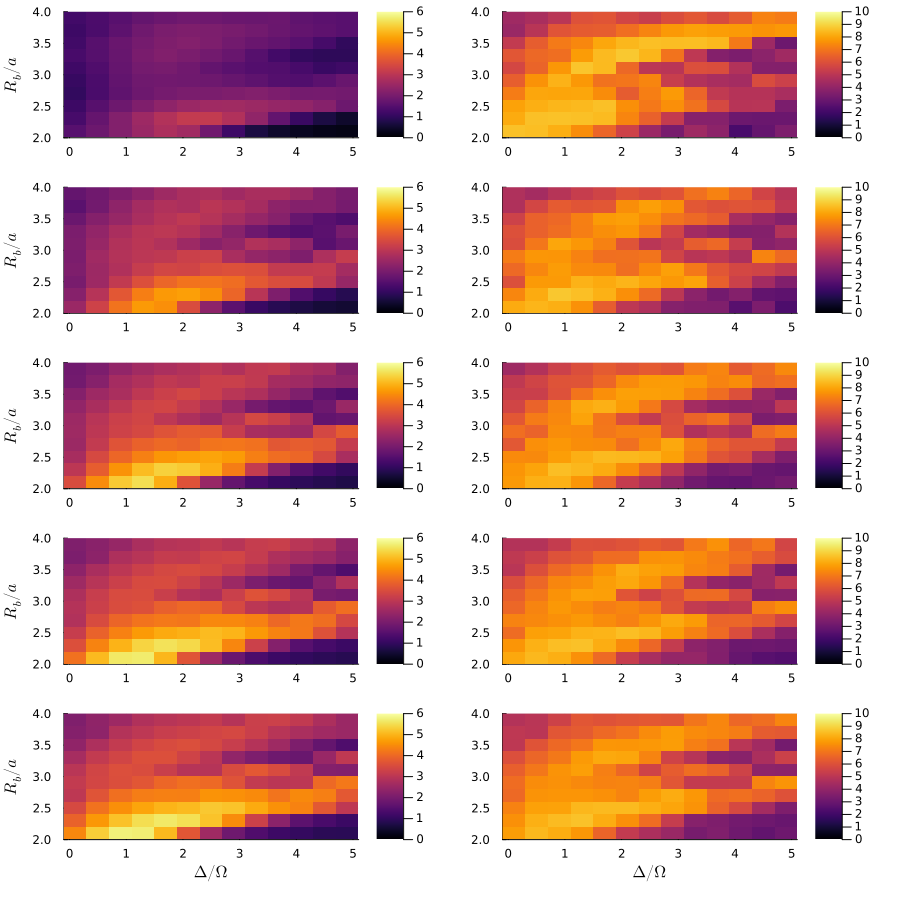}
    \caption{Half-rung entanglement entropy (left) and particle production (right) with time increasing down. $\Delta t=0.35 \; \mu s$. Results show a consistent dependence on model parameters ($R_b/a, \; \Delta/\Omega$).}
    \label{fig:movie}
\end{figure}

Figures \ref{fig:max_entropy}(a) and \ref{fig:hadrons} of the main text display single-time snapshots of the dependence on models parameters ($R_b/a, \; \Delta/\Omega$) of the half-rung entanglement entropy and average multiplicity, respectively. Figure \ref{fig:movie} shows each of these properties as a function of time to demonstrate the development of extended correlations and the proliferation of strings. These show a fairly consistent dependence on model parameters, demonstrating that particle production is highest in regions of the parameter space where entanglement grows quickly.

\onecolumngrid

\twocolumngrid

\bibliographystyle{apsrev4-1}
\bibliography{bibliography.bib}

\end{document}